**Title: CNN-based Survival Model for Pancreatic Ductal Adenocarcinoma in Medical Imaging**

**Authors:**

| # | Name | Affiliations |
|---|---|---|
| 1 | Yucheng Zhang | 1,2 |
| 2 | Edrise M. Lobo-Mueller | 3 |
| 3 | Paul Karanicolas | 4 |
| 4 | Steven Gallinger | 2 |
| 5 | Masoom A. Haider | 1,2,5 |
| 6 | Farzad Khalvati | 1,2 |

**Affiliations**

1: Department of Medical Imaging, Faculty of Medicine, University of Toronto, Toronto, ON, Canada

2: Lunenfeld-Tanenbaum Research Institute, Sinai Health System, Toronto, ON, Canada

3: Department of Radiology, Faculty of Health Sciences, McMaster University and Hamilton Health Sciences, Juravinski Hospital and Cancer Centre, Hamilton, Ontario, Canada

4: Department of Surgery, Sunnybrook Health Sciences Centre, Toronto, ON, Canada.

5: Sunnybrook Research Institute, Toronto, ON, Canada

**Correspnding Author:** Dr. Farzad Khalvati, M.A.Sc., Ph.D.

Staff Scientist, Lunenfeld-Tanenbaum Research Institute (LTRI), Sinai Health System

Assistant Professor, Medical Imaging, University of Toronto

Farzad.Khalvati@utoronto.ca

# Abstract

**Background:** Cox proportional hazard model (CPH) is commonly used in clinical research for survival analysis. In quantitative medical imaging (radiomics) studies, CPH plays an important role in feature reduction and modeling. However, the underlying linear assumption of CPH model limits the prognostic performance. In addition, the multicollinearity of radiomic features and multiple testing problem further impedes the CPH models performance. In this work, using transfer learning, a convolutional neural network (CNN) based survival model was built and tested on preoperative CT images of resectable Pancreatic Ductal Adenocarcinoma (PDAC) patients.

**Results:** The proposed CNN-based survival model outperformed the traditional CPH-based radiomics approach in terms of concordance index by 22%, providing a better fit for patients' survival patterns.

**Conclusions:** The proposed CNN-based survival model outperforms CPH-based radiomics pipeline in PDAC prognosis. This approach offers a better fit for survival patterns based on CT images and overcomes the limitations of conventional survival models.

## Background

As a statistical method, survival analysis is commonly used in clinical research to identify potential risk factors or biomarkers for a variety of clinical outcomes including patients' overall survivals for different diseases such as cancer. Cox proportional hazard model (CPH) is one of the most commonly used survival analysis tools[1–4]. CPH is a type of semiparametric model that calculates the effects of features (independent variables) on the risk of a certain event (e.g., death)[5]. For example, CPH can measure the effect of tumor size on the risk of death.

The CPH-based survival models can help clinicians make more personalized treatment decisions for individual patients. CPH models assume that the independent variables make a linear contribution to the model, with respect to time[5]. In many conditions, this assumption oversimplifies the relationships between biomarkers and outcomes, especially in cancer diseases with poor prognosis including Pancreatic Ductal Adenocarcinoma (PDAC)[3]. With a limited sample size, the violation of linear assumption may not be obvious. However, as data sizes increase, the violation of linear assumption in CPH models increasingly becomes more obvious and problematic, diminishing the performance and reliability of such models[2–4].

In real-world cases, nonlinear risk models can often provide a better fit for a survival function[3,6]. There are mainly two types of non-linear survival models: classification methods and risk-prediction methods[2–4]. Classification methods solve the nonlinearity by using a classifier such as Random Forest or Support Vector Machine (SVM)[7,8]. Although these classifiers perform well in many scenarios by providing binary predictions, they discard the duration information in modeling. For disease with poor prognosis such as pancreatic cancer, the 5-year survival rate is very low (e.g., less than 10% for pancreatic cancer)[9–11]. Consequently, binary predictions only offer limited information for healthcare professionals in designing personalized treatment plans and hence, a nonlinear survival model that takes duration (time to an event such as death) into account to provide useful information on the survival is desired.

Risk-prediction models which are based on artificial neural networks (ANNs) learn complex and nonlinear relationships between prognostic features and an individual's risk for a given outcome. Therefore, the ANNs-based model can provide an improved personalized recommendation based on the computed risk. Nevertheless, previous studies have demonstrated mixed performance for risk-prediction models[6,12,13]. This may be due to the small sample size and limited feature space leading to ANNs models that are underfitted[6]. To exploit the ANNs architecture and successfully apply them to complex cases, larger datasets are required. Recent work has shown that, given enough sample sizes, ANNs can, in fact, outperforms traditional CPH survival models[2–4].

In medical imaging, researchers have been working to extract diagnostic or prognostic features from medical images in different modalities[14–19]. Efforts have been made to standardize these quantitative imaging features (radiomic) by implementing open source libraries such as PyRadiomics[20]. These feature banks contain thousands of hand-crafted formulas, designed to extract the distribution or texture information. Subsequently, these features are often tested by CPH models selecting significant features and building the final survival model[21,22]. However, the high dimensionality nature of radiomics features introduces serious issues in feature reductions and prognosis performance through CPH models.

Through a standard radiomics feature bank, more than 1000 features can be extracted from each ROI. Given the high dimensionality of features, multiple testing in CPH models becomes a challenge[14]. In addition, the proposed feature sets are often highly correlated due to the similarity of formulas[16]. Despite the linear assumption in CPH modeling, the multicollinearity in the feature space further impedes the performance. The limitations of the handcrafted radiomic features and the fact that ANNs outperform traditional CPH models, motivate designing a novel approach for survival modeling that combines CPH with state-of-the-art deep learning algorithms for improved performance.

Previous work on deep learning based survival analysis including DeepSurv and NNET-survival are all ANNs-based survival models with modified loss function to capture more accurate survival patterns[2,3]. These models take features (e.g., age, gender, height) as input and return risks for patients at different timepoints. However, as discussed above, feeding radiomics features into these ANNs as input is not the optimal solution due to the multicollinearity issue. In this research, we use medical images as input, replacing conventional feature extractors with a Convolutional Neural Network (CNN) architecture to extract disease-specific image features which are associated with survival patterns. As the most well-known architecture in deep learning, CNNs extract imaging features by applying multiple layers of convolution operations to the images. Furthermore, the weights of the convolution filters are finetuned during training via backpropagation process[23,24]. Thus, given sufficient data, CNNs can be used to extract imaging features that are disease-specific, which can be used for diagnosis or prognosis purposes[25–28]. Although traditional medical imaging based CNNs use "binary" or "multinomial" classification loss function, the loss function can be modified to also capture the survival patterns[3]. By doing so, CNN can be tuned to extract features that are associated with the risk of the outcome in a certain duration. We hypothesized that CNNs would extract more meaningful features and the proposed CNN-based Survival (CNN-Survival) model with a modified loss function, which captures survival patterns would outperform conventional radiomics and CPH-based prognosis models.

## Methods

**Architecture of the proposed CNN-Survival**

A CNN architecture with six-layered convolutions (CNN-Survival) was trained as shown in Figure 1. Input images have dimensions of 140×140×1 (grey scale), which contain the CT images within the manual contours of the tumors (example shown in Figure 2). All convolutional layers have kernel size of 3×3 with 32 filters following by Batch Normalization layers (BN). The first Max Pool layer has pool size of 2×2, and the latter two Max Pool layers have pool size of 3×3. Through the Max Pool layers, number of trainable parameters was significantly reduced. To avoid overfitting with this small sample size, dropout layers were added after every two convolutional layers with dropout rate at 0.5. Finally, passing through the flatten layer, images were converted into 800 features where survival probabilities for a given time t were calculated.

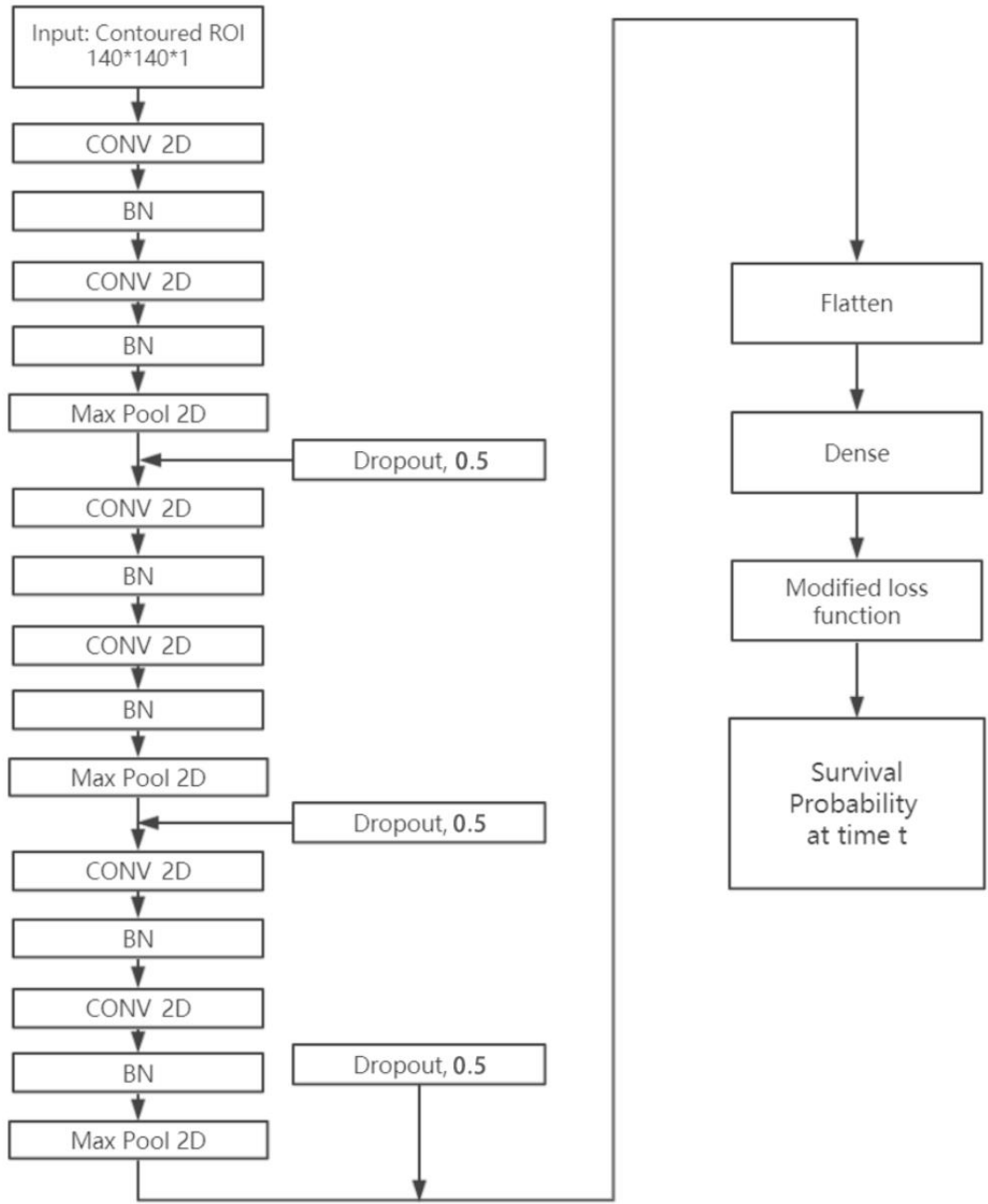

**Figure 1** The proposed CNN-Survival architecture: 6-layer CNN batch normalization and Max Pooling layers. There are also three dropout layers to control the potential overfitting.

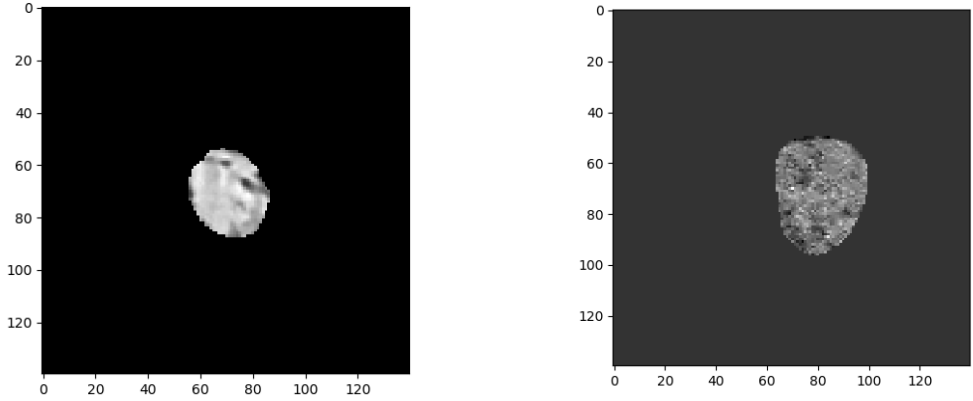

**Figure 2** Example of the input CT images

Left: NSCLC tumor from Cohort 1. Right: PDAC tumor from Cohort 2

**Loss Function**

To better fit the distribution of survival data, a modified loss function, proposed by Gensheimer et al[3], was applied to the CNNs architecture (Equation 1).

$$loss = -\sum_{i=1}^{d_j} \ln(h_j^i) - \sum_{i=d_j+1}^{r_j} \ln(1 - h_j^i) \qquad (1)$$

In the Equation 1, $h_j^i$ is the hazard probability for individual i during time interval j. $r$ stands for individuals "in view" during the interval j (i.e., survived in this period) and $d$ means a patient suffered a failure (e.g., death) during this interval[3]. As it can be seen from Equation 1, the left part penalizes if the model gave low hazard for failure (e.g., death), while the right part penalizes if the model gave high hazard for a survived case. The overall loss function is the sum of the losses for each time interval[3].

**Data**

CT scans along with patient outcome (survival and time to death) from three independent cohorts were used in this study. Cohort 1 consists of publicly available 422 Non-small cell lung cancer (NSCLC) patients[29]. Cohort 2 has 68 resectable pancreatic adenocarcinoma (PDAC) patients collected from a local hospital. Cohort 3, which is the test data, consists of 30 resectable PDAC patients enrolled in another independent hospital site[21]. For all the patients in these three independent cohorts, CT scans, annotations (contours) of tumor performed by radiologists, and survival data were available. For PDAC patients, the CT scans were preoperative images of resectable patients, and the survival data was collected from the date of surgery until death. The institutions' Research Ethics Boards approved these retrospective studies and all methods were carried out in accordance with relevant guidelines and regulations.

**Training process and Transfer Learning**

Training a CNN-based survival model needs to finetune a large number of features. Given this CNN architecture, there were 73,587 trainable parameters. As such, the larger dataset, cohort 1, was used to pretrain the network. In Cohort 2, 422 patients had 5,479 slices containing manually contoured tumor regions. However, the region of interest (ROI) on some of the slices were too small (e.g., less than 200 pixels) to be fed as input to the CNN (shown in Figure 3). To mitigate this, we ranked slices using their ROI size and pixel intensity and picked the top 2,500 slices. This ensured the minimum ROI size of 250 pixels.

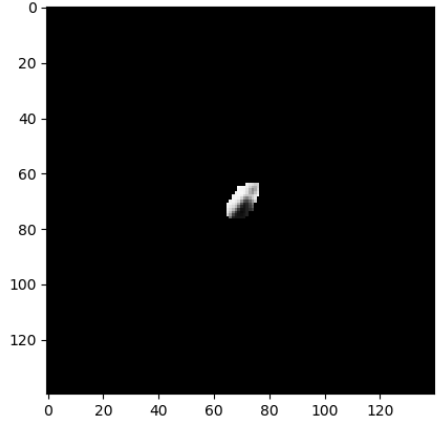

**Figure 3** Example of small ROI in Cohort 1

These 2,500 slices were fed into the proposed CNN model. After training the initial model for 50 epochs, all the weights in the pretrained model were frozen except for the final dense layer. Next, 68 patients from Cohort 2 were used to finetune the dense layer which contains 627 parameters. The finetuning was necessary since Cohort 1 and Cohort 2 have CT images from two different types of disease (lung and pancreas cancer, respectively) with different survival patterns. After 20 epochs of finetuning, the final model was tested in Cohort 3, and the concordance index at 1 year was calculated using Equation 2.

$$c = \frac{1}{|\mathcal{E}|} \sum_{T_i \text{ uncensored}} \sum_{T_j > T_i} \mathbf{1}_{f(x_i) < f(x_j)} \quad (2)$$

where the indicator function $1_{a<b} = 1$ if a < b, and 0 otherwise. $T_i$ is the survival time for subject i. $|\mathcal{E}|$ is the number of edges in the order graph. $f(x_i)$ is the predicted survival time for subject i by model f. Under this formula, concordance index (CI) is the probability of concordance between the predicted and the observed survival[30].

**Traditional Radiomics analytic pipeline**

To compare the prognostic performance of the proposed CNN-Survival model to traditional radiomic based CPH models, the following model was built. In Cohort 2 and Cohort 3, 2D radiomics features were extracted from the manually contoured regions using PyRadiomics library (version 2.0), generating 410 features in total[20,21]. A recent study using these two datasets has developed a radiomic signature that was shown to be associated with overall survival in resectable PDAC cohorts[21]. The formula of the signature is given in Equation 3.

$$Radiomics\ Signature = e^{0.44*F_1 + 0.11*F_2} \quad (3)$$

where $F_1$ is original_glcm_SumEntropy and $F_2$ is squareroot_glcm_ClusterTendency. In this study, a CPH model was trained using this Radiomics Signature in Cohort 2. Then, the performance of this radiomics based CPH model was validated in Cohort 3 by Concordance Index using R software (version 3.5.3) and SurvComp library[31].

## Results

Pretraining the proposed CNN-Survival with Cohort 1, given learning rate of 0.0001, the loss decreased significantly with the first ten epochs where the loss of training and validation sets converged quickly. Concordance index (CI) was used to measure the fit of the survival function. In Cohort 1 (training), CNN-Survival had a CI of 0.642. After finetuning using Cohort 2, the proposed model achieved a CI of 0.644 in Cohort 3 (test). However, if a CNN-Survival was trained from scratch in Cohort 2, the CI was 0.526 in the test cohort, indicating the importance of transfer learning in the small sample size setting. Previous radiomics study had identified a robust radiomics feature based signature across two different institutions[21]. However, in the test cohort, using CPH, the Radiomics Signature yielded CI of 0.528, which was significantly lower than that of the proposed CNN-Survival model as shown in Table 1.

**Table 1**: Results (concordance index) of three survival models for resectable PDAC

|  | Cohort 3 (test) |
|---|---|
| Radiomics Signature + CPH | 0.528 |
| Proposed CNN-Survival (Without Transfer Learning) | 0.526 |
| Proposed CNN-Survival (With Transfer Learning) | 0.644 |

As discussed above, CNN-Survival could depict the survival probability of a patient at a given time. The survival probabilities of two patients (one survived versus one deceased) in the test cohort are shown in Figure 4 and Figure 5.

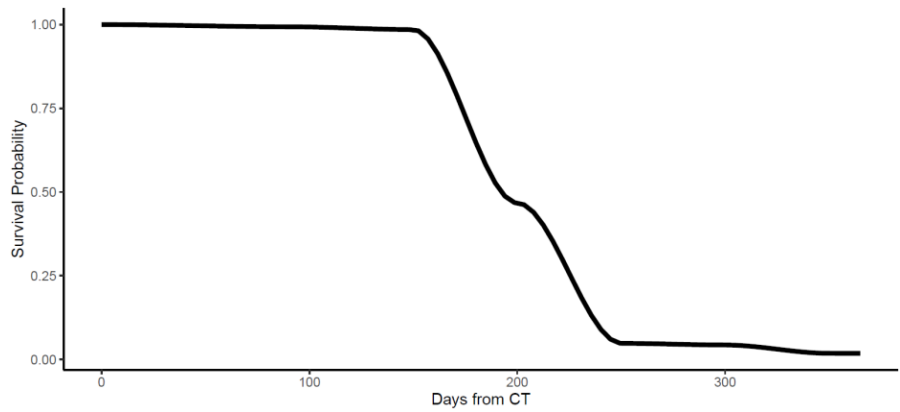

Figure 4. Survival probability curve generated by the proposed CNN-Survival model for a patient deceased 316 days after surgery.

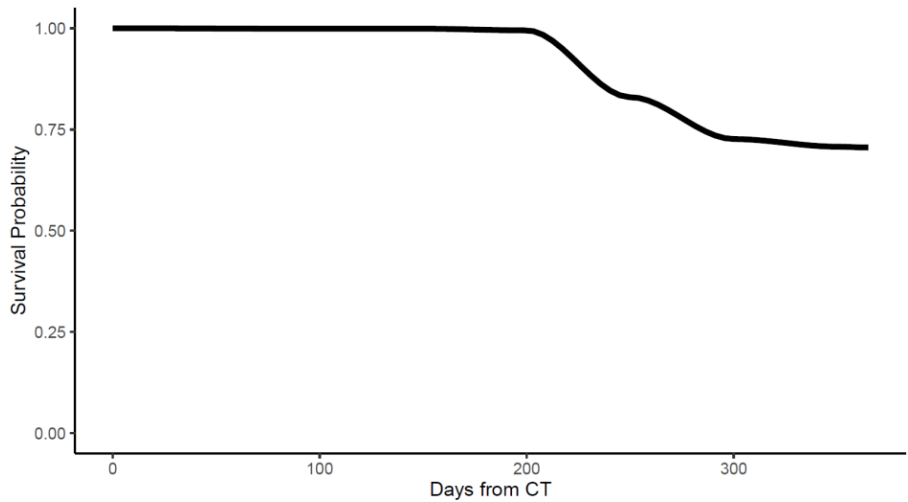

Figure 5. Survival probability curve generated by the proposed CNN-Survival for a patient survived more than one year after surgery.

For the patient deceased within one year after surgery, the survival probability dropped significantly, while for the survived patient, the survival probability stays above 0.5.

**Discussion**

Using the proposed CNN-Survival model, the prognosis performance was improved, elevating CI from 0.528 to 0.644. Deep learning networks provide flexibility in modifying the dimension of feature space and loss function, enabling us to extract disease-specific features and build more precise models. Using a CNN-based survival model, we showed that, with the help of transfer learning, deep learning architectures can outperform traditional pipeline in a typical small sample size setting in modeling survival for resectable PDAC patients. The proposed transfer learning-based CNN-Survival model has significant potential. For example, we can pretrain a model using images from common cancers with larger datasets and transfer this model to target rare cancers. Transfer learning-based CNN-Survival model mitigates the needs for large sample size, allowing the model to be applied to a wide range of cancer sites.

The proposed CNN-Survival model provides better performances compared to the traditional radiomics analytic pipeline. From the feature extraction perspective, parameters in a CNN can be updated during backpropagation, allowing to extract a large number of features that are associated with the target outcome. In contrast, given that the formula of each radiomics feature has been previously defined independent of outcome, some radiomics features may not be predictive for specific types of diseases. For feature analysis, the CNN-Survival model avoids the multiple testing, which is a significant issue in the conventional radiomics analytic pipeline. Finally, with the modified loss function, CNN-Survival model does not rely on the linear assumption, making it suitable for more real-world scenarios. These advantages contributed to the improved performance of the proposed model. In the test cohort, the proposed CNN-Survival model achieved a concordance index at 0.644. Although there was no prior publications reporting CI for PDAC patients, the CI of our proposed CNN-Survival is comparable to the typical CI for other biomedical survival models[32]. We also built a CPH model based on Radiomic Signature for resectable PDAC cases from a previous work[21], and demonstrated that the proposed CNN-Survival model outperforms the traditional radiomics-based survival models.

In this research, due to the small sample size in PDAC cohorts, the proposed CNN model was not optimal. We used CT images from 68 patients to finetune the pretrained CNN-Survival model and tested in another 30 patients of an independent cohort. Although through transfer learning, most of the parameters were trained using the pretrained cohort, there were still 627 parameters in the dense layer needed to be modified through finetuning. Thus, if a larger dataset was available for finetuning, performance may be further improved. Additionally, the pretrained dataset are CT images from Non-Small Cell Lung Cancer (NSCLC) patients. Although it is the largest open source dataset we could find, NSCLC has different biological traits and survival patterns compared to PDAC. In future research, using a similar pretrained domain and a larger finetuning cohort, further improvement may be achieved.

In this study, using CT images from three independent cohorts, we validated the proposed CNN-Survival with the modified loss function proposed by Gensheimer et al[3]. We showed that the proposed CNN-Survival model outperformed and avoided the limitations of the conventional radiomics-based CPH model in a real-world small sample size setting. Further validation of this loss function can be performed for other types of diseases through transfer learning. The proposed CNN-Survival model has the potential to be a standardized survival model in quantitative medical imaging research field.

# Conclusions

The proposed CNN-based survival model outperforms traditional CPH-based radiomics pipeline in PDAC prognosis. This approach offers a better fit for survival patterns based on CT images and overcomes the limitations of conventional survival models.

# Declarations

**Ethics approval and consent to participate**

Cohort 1 is publicly available and can be downloaded from: https://wiki.cancerimagingarchive.net/. For Cohort 2, University Health Network Research Ethics Boards approved the retrospective study and informed consent was obtained. For Cohort 3, the Sunnybrook Health Sciences Centre Research Ethics Boards approved the retrospective study and waived the requirement for informed consent.

**Authors' contributions**

YZ, MAH, and FK contributed to the design of the concept. EML, SG, MAH, and FK contributed in collecting and reviewing the data. YZ and FK contributed to the design and implementation of quantitative imaging feature extraction and machine learning modules. All authors contributed to the writing and reviewing of the paper. All authors read and approved the final manuscript. FK and MAH are co-senior authors for this manuscript.

**Competing interests**

The author(s) declare no competing interests.

**Data Availability**

The datasets of Cohort 2 and Cohort 3 analyzed during the current study are available from the corresponding author on reasonable request pending the approval of the institution(s) and trial/study investigators who contributed to the dataset.

**Funding Acknowledgment**

This study was conducted with the support of the Ontario Institute for Cancer Research (OICR, PanCuRx Translational Research Initiative) through funding provided by the Government of Ontario.